\documentstyle[preprint,eqsecnum,aps]{revtex}

\begin{document}
\begin{titlepage}
\begin{center}
{\Large\bf Metafluid dynamics as a gauge field theory}
\vskip.1in 
A. C. R. Mendes\footnote{Email: \tt albert@fisica.ufjf.br}$^a$, C.Neves\footnote{Email: 
\tt cneves@fisica.ufjf.br}$^b$, W. Oliveira\footnote{Email: \tt wilson@fisica.ufjf.br}$^b$, 
and F.I. Takakura\footnote{Email: \tt takakura@fisica.ufjf.br}$^b$\\~\\ 
$^a$ Centro Brasileiro  de Pesquisas F\'\i sicas, Rio de Janeiro, 
22290-180, RJ, Brasil,\\

$^b$ Departamento de F\'\i sica, ICE, Universidade Federal de Juiz de Fora,\\
36036-330, Juiz de Fora, MG, Brasil
\end{center}

\begin{abstract}
In this paper, the analog of Maxwell electromagnetism for hydrodynamic turbulence, the metafluid dynamics, is extended in
order to reformulate the metafluid dynamics as a gauge field theory. That analogy opens up the possibility to investigate this theory as a constrained system. Having this possibility in mind, we propose a Lagrangian to describe this new theory of turbulence and, subsequently, analyze it from the symplectic point of view. From this analysis, a hidden gauge symmetry is revealed, providing a clear interpretation and meaning of the physics behind the metafluid theory. Further, the geometrical interpretation to the gauge symmetries is discussed and the spectrum for $3D$ turbulence computed.

\end{abstract}
\leftline{PACS numbers: 47.27.-i, 11.90.+t}
\leftline{Keywords: Constrained Systems, Gauge Theory, Turbulence}
\end{titlepage}

%\pacs{PACS numbers: 47.27.-i, 11.90.+t} 

%\newpage 
\section{Introduction} 

The understanding of hydrodynamic turbulence is an important problem for nature science, from both, theoretical and experimental point of view,
and has been investigated intensively \cite{Landau,Monin-Yagolm,Frisch} over the last century, but a deep and fully comprehension of the problem remains obscure.

Over the last years, the investigation of turbulent hydrodynamics has experienced a revival since turbulence has became a very fruitful
research field for theoreticians, that study the analogies between turbulence and field theory, critical phenomena and condensed matter
physics \cite{Bramwell-Holdsworth-Pinton,L'vov,Periwal,Polyakov,Gurarie,Eyink-Goldenfeld,Nelkin,deGennes,Rose-Sulem}, renewing the optimism to solve the turbulence problem.

The dynamics of turbulent viscous fluid is expressed by the Navier-Stokes (NS) equations of motion\cite{Landau}, which in vectorial
form is

\begin{equation}
{\partial \vec {\rm u} \over{\partial t}} = - \vec {\rm w} \times \vec {\rm u} - 
\nabla \left({p\over \rho} + {{\rm u}^2\over 2}\right) + \nu \nabla^2 \vec {\rm u}, \label{01}
\end{equation}
where $\vec {\rm u} (\vec  x, t)\;$ is the velocity field, $\vec {\rm w} (\vec x, t)\;$ 
the vorticity field, $p(\vec x, t)\;$ is the pressure, $\rho\;$ the  density and $\nu\,
$ the kinematic viscosity.

The equation of continuity reduces to the requirement that the velocity field is 
divergenceless for incompressible fluids, i.e, 
\begin{equation}
\label{02}
\nabla .\vec {\rm u} =0,
\end{equation}
which are the flows we are interested in this paper. In this context, the hydrodynamic turbulence has attracted an enormous
interest due to the universal characteristics stressed by an incompressible fluid with high Reynolds numbers in the fully developed
turbulent regime. The Reynolds number, $R \equiv LU/\nu $ (where  $L\;$ is the integral 
length-scale of the largest eddies and $U$ is a characteristic large-scale velocity), measures the competition between convective and
diffusive processes in 
an incompressible fluid described by the NS equations. In view of this, the incompressible fluid flow assumes high Reynolds numbers when
the velocity increases and, consequently, the solution for eqn.(\ref{01}) becomes unstable and the fluid switches to a new regime of a very
complex motion with the velocity varying almost randomly and without any noticeable order. To discover the laws describing what
exactly is going on with the fluid in this turbulent regime is very important to both theoretical and applied science.

Recently, Marmanis\cite{Marmanis} has proposed an alternative approach to treat fluid turbulence. Based on the analogy between Maxwell
electromagnetism and turbulent hydrodynamics, he describes the dynamical behavior of average flow quantities in incompressible fluid flows
with high Reynolds numbers. In this theory, metafluid dynamics, the vorticity 
$(\vec {\rm w}=\nabla \times \vec {\rm u})$ and Lamb vector $(\vec {\rm l} = \vec {\rm w}\times 
\vec {\rm u})$ are recognized as the kernel of this dynamical theory of turbulence. In our paper, we make a further investigation of the physical contents present in that theory, in order to furnish a better understanding of turbulence.

To extract the physical meaning of the metafluid dynamics, we extend the Marmanis analogy in order to propose an appropriate Lagrangian governing the dynamics of incompressible fluid flows. From this Lagrangian and using the symplectic method\cite{FJ}, the physics behind the metafluid theory is discovered. It is a new and strike result not yet discussed in the literature.

In order to make this work self-consistent, we have organized this paper as follows. In section 2, a brief review of the metafluid
dynamics is presented as well as the pertinent physical quantities and notation. In section 3, the Lagrangian approach for the metafluid
dynamics is proposed and some considerations about the dissipation rate of energy is discussed. In section 4, the new
description for the metafluid dynamics is quantized using the symplectic method, which leads to unveil the gauge symmetry.
Further, we scrutinize the symmetry of the model and show that the gauge invariance is only preserved in some limits,
called {\it inertial range}, and also we identify the physical quantities that are gauge invariant into this region.
In section 5, the geometric interpretation of the gauge symmetry is given and discussed. In section 6, the spectrum of the metafluid
theory is computed and the result analyzed and compared with the usual results in the literature\cite{spectrum}. In Sec.7, we stress our conclusion and final discussions. We added an appendix with a brief review of the Simplectic formalism.

\section{Metafluid dynamics} 

The problem of turbulence is to find the averaged properties from the solutions of NS equations under the constraint of the
incompressibility condition, eqn.(\ref{02}), forming a system of coupled nonlinear partial differential equations. Due to this, it is
a difficult task to get a common feature of averaged nonlinear equations, because nonlinearity introduces higher order momenta of
fluctuation, what is known as closure problem. To overcome this problem, Marmanis\cite{Marmanis} proposed an approximate theory
of turbulence based on the analogy between Maxwell electromagnetism and turbulent hydrodynamics, where the equations governing the dynamic
variables become linear and the nonlinearities appear as sources of turbulent motion. In this picture, one constructs a system of equations written
in terms of the average values of the vorticity $(\vec {\rm w})$ and Lamb vector $(\vec {\rm l})$, instead of the average values of
$\vec {\rm u}$ and $p$. Essentially, it  turns the turbulence closure problem into a study of turbulent sources for different
geometries of interest.

In this scenario, the study of average quantities of an incompressible fluid at the 
fully developed regime is proposed, given place to a system where the average fields show up in a continuum inter-relation and
respond as waves to the turbulent sources. To do so, the Lamb vector and the vorticity should be taken as the kernel of
the turbulent dynamics rather through velocity and vorticity fields or velocity and pressure fields. Then whatever parts that can not be explicitly expressed as a function of
$\vec {\rm w}$ or $\vec {\rm l}$ only, are gathered and treated as source terms. This is done by introducing the concepts of
{\it turbulent charge} $(\rm n)$ and {\it turbulent current} $(\vec J)$. The turbulent charge is connected with the Bernoulli energy
function,
\begin{equation}
\label{03} 
\Phi({\vec x},t)=\frac{p}{\rho}+\frac{{\rm u}^2}{2},
\end{equation}
through the relation
\begin{equation}
\label{03a} 
{\rm n}(\vec x, t)= - \nabla ^{2} \Phi.
\end{equation}

In this formalism, the equations of motion describing the behavior of the hydrodynamic turbulence are
\begin{eqnarray}
{\nabla .\vec  w} &=& 0, \nonumber \\
{{\partial \vec w} \over {\partial t}} &=& -{\nabla} 
\times {\vec l} + \nu {\nabla }^{2}{\vec w}, \nonumber \\ 
\nabla .\vec l &=& n({\vec x}, t), \label{04} \\
{{\partial \vec l} \over {\partial t}} &=& c ^{2}\nabla \times \vec w - 
{\vec J}({\vec x}, t) + \nu {\nabla }n({\vec x}, t) - \nu {\nabla } 
^{2}{\vec l }, \nonumber
\end{eqnarray}
where $c^2 =\langle {\rm u}^2 \rangle$  is the spatial averaged squared velocity, $\vec w$ and $\vec l$ are defined as the averages of $\vec {\rm l}$ and $\vec {\rm w}$, while the sources $\vec J$ and $n$ are averages of $\vec j$ 
and ${\rm n}$, respectively. Further, the Lamb vector can be written in terms of the velocity field and the Bernoulli energy function from the NS
equations as
\begin{equation}
\vec l = -{{\partial \vec u} \over {\partial t}} -{\nabla }\phi + 
\nu {\nabla }^{2} { \vec u}, \label{05}
\end{equation}
where $\vec u =\langle \vec {\rm u}\rangle$ and $\phi =\langle \Phi \rangle$.

Now, we would like to compare the set of eqn.(\ref{04}) with the Maxwell equations with sources in vacuum\cite{Jackson}. Note that
these set of equations are identical if $\vec B$ (magnetic field) corresponds to $\vec w$ and $\vec E$ (electric field) corresponds to
$\vec l$. Furthermore, the analogy can be extended in order to include the potentials as 
well. In particular, the comparison suggests that the vector potential corresponds to the 
velocity field $\vec u$ and the scalar potential to the Bernoulli energy function $\phi$.  

Despite of all resemblance between hydrodynamic turbulence and electromagnetic 
theory, there is a conceptual difference in the identification of the physical entities. In the 
classical electromagnetism, the physical fields are the electric and magnetic fields while the 
potentials are just mathematical artifices. Oppositely, in the metafluid dynamics, the potentials are the entities which have physical
significance.

\section{Lagrangian approach}

From the geometric point of view and using the Lie
algebra, Arnold \cite{Arnold} showed that Euler flow can be described in the Hamiltonian formalism in any dimension. This has a lot of
interesting consequences for fluid mechanics and has been studied intensively\cite{Arnold-Khesin,Zeitlin}. However, it is not quite
obvious that this process can be used when viscosity is taken in account. It is in this scenario where the metafluid dynamics births,
revealing a way to find a Hamiltonian formalism for a turbulent flow with viscosity.

In the classical electromagnetism, the Lagrangian density can be written as the difference between the square of the electric and magnetic fields, as follows,

\begin{equation}
{\cal L} = {1\over 2} (\vec E^2 - \vec B^2).\nonumber
\label{06}
\end{equation}
Using the analogy established between electromagnetism and turbulence, one can write down the Lagrangian density for the theory of
turbulence as

\begin{equation} 
\label{07} 
{\cal L} ={1 \over 2} ({\vec l}\,^{2} -c^{2}{\vec w} ^{2}), 
\end{equation}
that can also be written in terms of velocity field and the Bernoulli energy function, named ``potentials of theory'', as  
\begin{equation} 
\label{08} 
{\cal L} ={1 \over 2}\left( -{\nabla }\phi -{\partial {\vec u} \over 
{\partial t}} + {\nu } {\nabla } ^{2}{\vec u} \right) ^{2} -{1 \over 2}c^{2} 
( \nabla \times {\vec u}) ^{2} . 
\end{equation} 
It is easy to see that this Lagrangian density gives us the equations of motion (\ref{04}) for the homogeneous case (no sources).

Now, let us consider the case when sources do not vanish. In this case, the sources appear in the equations of motion and, as  a consequence, an interaction Lagrangian, defined as
\begin{equation} 
\label{09} 
{\cal L}_{int} ={\vec J}.{\vec u} - n\phi 
 -{\nu }{\vec u} .{\nabla }n, 
\end{equation} 
which contains a viscous correction term, is added to the Lagrangian (\ref{08}), providing the total Lagrangian density,
\begin{eqnarray} 
\label{10} 
{\cal L} = {1 \over 2} \left( -{\nabla }\phi -{{\partial \vec u} \over {\partial t}} + {\nu }{\nabla } ^{2}{\vec u} \right) ^{2} -{1 \over 2} c^{2}({\nabla } \times
{\vec u}) ^{2}+  {\vec J} .{\vec u}
-n\phi  - {\nu }{\vec u} .{\nabla }n. 
\end{eqnarray} 
There is no doubt that the metafluid dynamics can produce some interesting results. Hence, let us show that the equations of motion for
viscous fluid, given in eqn.(\ref{04}), can be found from the Euler-Lagrange equations, where the velocity field $(\vec u)$ and the
Bernoulli energy function $(\phi)$ are the canonical variables. From the Lagrangian density, eqn.(\ref{10}), NS equations
(\ref{01}) will be obtained in order to demonstrate that the physical contents of turbulence in the NS equations description are also present in our
construction. The conjugated momenta of velocity field is computed as
\begin{equation}
\label{11}
\vec \pi(\vec x,\,t) ={{\delta {\cal L}}\over{\delta \dot {\vec u}}(\vec x,\,t)}=
{{\partial\vec u}\over {\partial t}} +\nabla\phi -\nu \nabla^2 \vec u=
-\vec l(\vec x,t),
\end{equation}
and using (\ref{05}), we get the NS equations, namely,
\begin{equation}
\label{12}
{{\partial}\over {\partial t}}\vec u(\vec x,\,t)=-\vec l(\vec x,t) -\nabla\phi(\vec x,t) 
+\nu \nabla^2 \vec u(\vec x,t),
\end{equation}
where $\,\vec l(\vec x,\,t)=\vec w \times \vec u$.

One of the simplest consequences of the NS equations (\ref{01}), modified by the addition of 
the external force $\vec f$, may be obtained by taking its scalar product with $\vec 
{\rm u}$ and integrating the result over the space coordinate. The result is the relation
\begin{equation} 
\label{13} 
{{\rm d} \over {\rm d}t}\int {1 \over 2}{\rm u}^{2} = -{\nu \over 2}
\int ({\nabla \vec {\rm u}})^{2} 
+\int {\vec f.\vec {\rm u}}, 
\end{equation} 
which expresses the energy balance: the time derivative of  energy on the right hand side
 is equal to the difference of the injection rate $\int \vec f.\vec {\rm u}$ and the dissipation of energy
 rate ${\nu \over 2} \int {\left(\nabla \vec {\rm u} \right)}^2$. Taking averages 
in the stationary state, we obtain
\begin{equation}
\label{14}
\langle \vec {\rm u} .\vec f \rangle = \langle {\nu \over 2}(\nabla \vec {\rm u})^2 \rangle 
\equiv \epsilon , 
\end{equation}
that is the equality of the (intensive) mean injection and the mean dissipation rates of energy. 
The energy injection takes place at the distances of order of the integral scale by induction of
 big scale $L$ eddies. According to the picture of the turbulent flow proposed in 1922 by 
Richardson \cite{Richardson}, the big eddies induce still smaller eddies and so on transferring
 energy from large to small distance scales. This process should not lead to a loss of energy 
until sufficiently small distance scales, say, smaller than $\eta$ (dissipative scales of 
Kolmogorov\cite{Frisch}), are  reached. On scales smaller than $\eta$, the dissipative term 
$\nu \nabla^2 \vec {\rm u}$  of the NS equation becomes important.

Considering now the intermediate domain of scales called the {\it inertial range}:
\begin{center}
$\eta \ll$ Inertial Range $\ll L$,
\end{center}
where the injection rate and the dissipation rate of energy should be negligeable, the Lagrangian density for the metafluid dynamics becomes
\begin{eqnarray} 
\label{15} 
{\cal L} =  {1 \over 2}\left( -{\nabla}\phi -{{\partial \vec u} \over {\partial 
t}}\right)^{2} - {1 \over 2}c^{2}(\nabla \times {\vec u})^{2} +   
{\vec J}.{\vec u} - n\phi . 
\end{eqnarray}

From this Lagrangian, the analogy between hydrodynamic turbulence and the electromagnetism will be extended. Note that in eqn.(\ref{05}),
the fields $\vec l$ and  $\vec w $, as happens with the electric and magnetic fields, are components of an antisymmetric second rank tensor,
the strength tensor, defined as
\begin{equation} 
\label{16} 
F^{\mu \nu} =\partial^{\mu}V^{\nu} -\partial^{\nu}V^{\mu}, 
\end{equation}            
which can be written out explicitly for $\vec l $ and $\vec w $  as
\begin{equation} 
\label{17} 
F^{\mu \nu} =\left(\begin{array}{cccc} 
0 & -l_x & -l_y & -l_z \\ 
l_x & 0 & -w_z & w_y \\ 
l_y & w_z & 0 & -w_x \\ 
l_z & -w_y & w_x & 0 
\end{array}\right) .
\end{equation} 
The four-vector potential and four-vector current are 
\begin{eqnarray} 
\label{18} 
V ^{\nu } &=& (\phi, c{\vec \emph u}), \nonumber\\
J^\nu &=& (cn, {\vec J}),
\end{eqnarray} 
respectively. Therefore, one can write (\ref{15}) in term of the $F^{\mu \nu}$, $J^\mu $ and $V^\mu$ as 
\begin{equation} 
\label{20} 
{\cal L} = -{1 \over 4}F_{\mu \nu}F^{\mu \nu} - {1\over c}J_{\mu}V^{\mu}. 
\end{equation}
It is easily verified through the Euler-Lagrange equations that (\ref{20}) gives the 
equations of motion for the inhomogeneous metafluid dynamics (\ref{04}) with $\nu =0$. The homogeneous equations of motion are obtained
from the dual of $F^{\mu \nu}$, given by
\begin{equation} 
\label{21} 
{\tilde F}^{\mu \nu } = {1 \over 2}\epsilon ^{\mu \nu \alpha \beta 
}F_{\alpha \beta} .
\end{equation} 

\section{Symplectic analysis}

In the last sections, the analogy between turbulent hydrodynamics and Maxwell electromagnetism was explored and a gauge field theory to
describe the turbulent fluid, called metafluid theory, was proposed, suggesting that this theory can be analyzed as a constrained
system. Both systems are characterized in phase space by the presence of some functions that depend on the coordinates and canonical momenta,
denominated constraints, which restrain the dynamics of the model. There are some methods to handle constrained systems, however we analyze the metafluid theory from the symplectic point of view\cite{FJ}. In the appendix we present the symplectic formalism for completeness.

To implement the symplectic formalism, the Lagrangian density (\ref{10}) will be rewritten in its first-order form, given by, 
\begin{equation}
\label{31}
{\cal L}^{(0)} =\vec \pi. \dot{\vec u} - U^{(0)},
\end{equation}
where the canonical momenta and the zeroth-iterative symplectic potential $(U^{(0)})$ are
\begin{eqnarray}
\label{32}
\vec\pi&=&  \nabla\phi + \dot{\vec u} - \nu \nabla^2\vec u,\nonumber\\
U^{(0)} &=& {1\over 2} \vec \pi^2 - \vec \pi . \nabla \phi - {1\over 2} c^2(\nabla \times \vec u)^2 + \nu \vec \pi . \nabla^2 \vec u - \vec u .
\vec J + \phi n + \nu \vec u . \nabla n,
\end{eqnarray}
respectively. 

From the set of symplectic variables $\xi_i^{(0)} = (u_i, \, \pi_i, \, \phi)$ and their respective one-form canonical momenta,
\begin{eqnarray}
a^{(0)}_{u_i} &=& \pi_i,\nonumber\\
a^{(0)}_{\pi_i} &= &0, \\
a^{(0)}_\phi &=& 0,\nonumber
\label{33}
\end{eqnarray}
the symplectic matrix is computed as
\begin{equation}
\label{34}
f^{(0)}= \pmatrix{ 0 & -\delta_{ij} & 0 \cr \delta_{ij} & 0 & 0\cr 0 & 0 & 0}\delta(\vec x - \vec y),
\end{equation}
which is singular, so, has a zero-mode $(\tilde v^{(0)} = (v^\phi , \, 0, \, 0))$. Contracting this zero-mode with the gradient of
symplectic potential $U^{(0)}$, the following constraint appears,
\begin{equation}
\label{35}
\Omega_1 =  \nabla . \vec \pi (\vec x) + n (\vec x).
\end{equation}
In agreement with the symplectic formalism, this constraint is introduced into the Lagrangian through a Lagrange multiplier $\lambda$,
namely,
\begin{equation}
\label{39}
{\cal L}^{(1)} =  \vec \pi .\dot {\vec u}  + \dot \lambda \Omega_1 - U^{(1)}
\end{equation}
where the symplectic potential density is
\begin{equation}
\label{40}
U^{(1)} = U^{(0)}\mid_{\Omega_1=0} = {1\over 2} \vec \pi^2 - {1\over 2}c^2(\nabla \times \vec u)^2 + \nu \vec \pi . \nabla^2 \vec u -
\vec u . \vec J + \nu \vec u . \nabla n.
\end{equation}

Considering now that the new set of symplectic variables is given in the following order, $\xi^{(1)}_{i} = (u_i,\, \pi_i, \, \lambda)$,
we have the one-form canonical momenta as,
\begin{eqnarray}
\label{41}
a^{(1)}_{u_i} &=& \pi_i ,\nonumber \\
a^{(1)}_{\pi_i}&=&0, \\ 
a^{(1)}_\lambda &=& \nabla . \vec \pi + n .\nonumber
\end{eqnarray}

The symplectic matrix $f^{(1)}$ is 
 \begin{equation}
\label{42}
f^{(1)}= \pmatrix{ 0 & -\delta_{ij} & 0 \cr \delta_{ij} & 0 & \partial_j^x \cr 0 & - \partial_i^y  & 0}\delta(\vec x - \vec y),
\end{equation}
that is a singular matrix with a zero-mode given by, 
\begin{equation}
\label{42.1}
\tilde v^{(1)} = (v^{\vec u}_{j},\, 0, \, v^\lambda),
\end{equation}
satisfying the following relation,
\begin{equation}
\label{43}
v_i^{\vec u} - \partial_i v^\lambda = 0. 
\end{equation}

Contracting the zero-mode $(\tilde v^{(1)})$ with the gradient of the symplectic potential  $(U^{(1)})$, we get a new constraint,
\begin{eqnarray}
\label{44}
\Omega_2 &=&\int {\rm d}^{3}{\vec x}\,v^{\vec u}_{i}{\delta \over{\delta u_{i}(\vec x,\,t)}}\int{\rm d}^3 {\vec y} \lbrace -
{1\over 2}c^2 (\nabla \times \vec u)^{2} +\nu \vec \pi .\nabla^2 \vec u +\nu \vec u .\nabla n -\vec u .\vec J \rbrace \nonumber \\
&=& \int {\rm d}^{3}\vec x \,v^{\vec u}_{i}(\vec x)J_{i}(\vec x).
\end{eqnarray}

\noindent Using (\ref{43}), $\Omega_2$ becomes
\begin{equation}
\label{45}
\Omega_2 = \int {\rm d}^{3}{\vec x}\,v^\lambda(\vec x)\partial_{i}J_{i}(\vec x) =
-\int {\rm d}^{3}{\vec x}\,v^\lambda(\vec x)\frac{\partial}{\partial t}\nabla^{2}\phi(\vec x).
\end{equation}

\noindent At this point, it is important to regard that at the inertial range, despite of the existence of viscosity, there is not energy
dissipation, then, the Bernoulli energy function $(\phi)$ is constant. Due to this, $\Omega_2$ vanishes and, consequently, the hidden gauge
symmetry of the metafluid theory is revealed.

To finish the symmetry analysis, it is necessary to obtain the infinitesimal gauge transformation. In the symplectic context, the gauge
transformations are generated by the zero-mode that does not produce a new constraint. In the present case, the zero-mode $\tilde v^{(1)}$,
given by eqn.(\ref{42.1}), does not generate a new constraint, consequently, it is the generator of the following infinitesimal gauge
transformations, 
\begin{eqnarray}
\label{54}
\delta u_{i} &=& \partial_{i}\epsilon, \nonumber \\
\delta \pi_{i} &=& 0, \\
\delta \phi &=& - \dot\epsilon,\nonumber
\end{eqnarray}
where  $\phi\rightarrow\dot\lambda$ and $\epsilon$ is an infinitesimal time-dependent parameter. It is easy to verify that the Lagrangian
(\ref{20}) is invariant under these transformations, 
\begin{eqnarray}
\delta {\cal L} &=& - ( \partial_i \pi_i + n)\dot\epsilon + ( - \dot n + \dot n) \epsilon + (\partial_i \pi_i + n) \nu \partial^2\epsilon  \nonumber \\
 &=& 0.
\end{eqnarray}
However, for the Hamiltonian below,
\begin{equation}
\label{55}
{\cal H} = {1\over 2} \vec \pi^2 - \vec \pi . \nabla \phi - {1\over 2} c^2(\nabla \times \vec u)^2 + \nu \vec \pi . \nabla^2 \vec u - \vec u .
\vec J + \phi n + \nu \vec u . \nabla n,
\end{equation}
we get
\begin{equation}
\label{56}
\delta{\cal H}= \epsilon \partial_{i}J_{i}=-\epsilon \nabla^{2}{\partial\over{\partial t}}\phi.
\end{equation}
We can observe that the gauge invariance is only preserved at the inertial range, where the Bernoulli energy function is constant.

In order to obtain the Dirac brackets among the phase space fields, we have to fix the gauge symmetry. It is done introducing a gauge
condiction into the kinetical sector of first-order Lagrangian through a Lagrange multiplier. So, we choose a gauge
fixing term that satisfies the condition of incompressibility of fluid, namely,
\begin{equation}
\chi=\nabla. \vec u.
\end{equation}
In view of this, the twice iterated Lagrangian is obtained as
\begin{equation}
\label{46}
{\cal L}^{(2)} =  \vec \pi .\dot{\vec u}  + \dot \lambda \Omega_1 + \dot\eta\chi - U^{(2)},
\end{equation}
where the symplectic potential density is
\begin{equation}
\label{47}
U^{(2)} = U^{(1)}\mid_{\chi=0} = {1\over 2} \vec \pi^2 + c^2\vec u . \nabla^{2}\vec u + \nu \vec \pi . \nabla^2 \vec u - \vec u . \vec J .
\end{equation}
Considering now that the new set of symplectic variables is given in the following order
$\xi^{(2)}_i = ( u_{i},\,\pi_{i},\,\lambda,\,\eta)$, we have
\begin{eqnarray}
\label{48}
a^{(2)}_{u_i} &=& \pi_i ,\nonumber \\
a^{(2)}_{\pi_i} &=& 0, \\ 
a^{(2)}_{\lambda} &=& \nabla . \vec \pi + n,\nonumber\\
a^{(2)}_{\eta} &=& \nabla . \vec u .\nonumber
\end{eqnarray}

\noindent Then, the corresponding symplectic matrix is obtained as
\begin{equation}
f^{(2)} = \left( \begin{array}{cccc}
                 0 & -\delta_{ij} & 0 & -\partial_i^x \\
                 \delta_{ij} & 0 & -\partial_i^x & 0\\
                 0 & - \partial_j^x & 0 & 0\\
                 - \partial_j^x & 0 & 0 & 0 \\ 
                 \end{array}
          \right)\delta(\vec x - \vec y). \label{49}
\end{equation}
This matrix is nonsingular and, consequently, the corresponding inverse matrix can be determined after a straightforward calculation. From
the inverse of $f^{(2)}\,$, the nonvanishing Dirac brackets among the phase space fields is automatically identified, namely,
\begin{equation}
\label{51}
\{u_i(\vec x),\, \pi_j(\vec y)\}^* = \left( \delta_{ij} - {\partial_i^x\partial_j^x\over{\nabla^2}}\right)\delta(\vec x - \vec y).
\end{equation}

The next step would be the quantization of this constrained theory. Using the well known canonical
quantization rule $(\{\;, \;\}^* \rightarrow -i [\;, \;])$, we have,
\begin{eqnarray}
\left[ u_i(\vec x), u_j(\vec y)\right] = \left[\pi_i(\vec x), \, \pi_j(\vec y)\right] &=& 0 , \nonumber \\
\left[ u_i(\vec x),\pi_j(\vec y)\right] &=& - i \left( \delta_{ij} - {\partial_i^x 
\partial_j^x\over{\nabla^2}}\right) \delta(\vec x - \vec y).\label{52}
\end{eqnarray}

Once we have the canonical quantization rule, we can apply standard Quantum Field Theory techniques to find the generating functional and,
consequently, the correlation functions and all physical quantities \cite{Itzykson,Henneaux} one wish.

\section{Geometric aspect of the gauge symmetry}

In this section, we give a geometrical interpretation to the gauge symmetry present on the metafluid theory (gauge field theory). The description of fundamental particle interaction, with the assistance of the gauge field theory, introduces extra degrees of freedom into the theory, that manifests itself in the singular nature of the respective Lagrangian or the presence of first class constraint in the equivalent Hamiltonian formulation. In this case, the phase space is larger than the physical one, which is a hypersurface determined by the constraints of the theory. In this gauge invariant scenario, the gauge potentials form an overcomplete basis and the gauge fields, which can be connected by an infinitesimal transformation, describe the same physical state. Thus, the gauge potentials are separated into equivalence classes with respect to the gauge group action, where each one denotes an orbit in the gauge field configuration. Transitions along(vertical) the orbits correspond to pure gauge transformations, consequently, these paths have no physical significance. Oppositely, perpendicular(horizontal) paths to the orbits describe the time evolution of the physical system, then, they are physically important to the theory. Hence, to find out the equations of motion for the physical fields consist in solving the problem of constructing the horizontal paths, which is computed just doing a correct definition of the metric in the orbit space. To do so, we follow process developed in \cite{NILL}, where the authors demonstrated that the physical (orbit) space is equipped with a natural projective metric.

In view of this, our task is to get a singular Lagrangian, dynamically equivalent to (\ref{20}), to govern the metafluid theory. It is achieved eliminating the Bernoulli energy function $(\phi)$ from the Lagrangian (\ref{20}). It can be done since this field has no dynamics, as demonstrated in the last section. To eliminate this field from (\ref{20}), we use the Euler-Lagrange equation of motion for $\phi$, given by,
\begin{equation}
\label{75}
\phi(\vec x)=-{{\partial_{i}\dot {u}_{i}+n}\over {\partial^{2}}}+\nu \partial_{i}u_{i}.
\end{equation}

\noindent Bringing back this result into the Lagrangian (\ref{20}) and rewriting it in terms of field components, we get
\begin{eqnarray}
\label{78}
{\cal L} &=& {1\over2}\dot{u}_{i}M_{ij}\dot{u}_{j}+(\partial_{i}\dot{u}_{i}){n\over\partial^{2}} +
{1\over2}{n^2 \over\partial^{2}}-\nu(\partial^{2}\dot{u}_{i})M_{ij}u_{j}\nonumber \\
&+&{\nu^{2}\over2}(\partial^{4}u_{i})M_{ij}u_{j}+{1\over2}c^2 ({\partial^{2}u_{i}})M_{ij}u_{j} +
u_{i}J_{i},
\end{eqnarray}
where the  metric of configuration space $M_{ij}$, reads as
\begin{equation}
\label{79}
M_{ij}=\delta_{ij} - {{\partial_{i}\partial_{j}}\over{\partial^{2}}},
\end{equation}
being a singular matrix which has $\partial_{i}$ as eigenvectors with zero eigenvalues,
\begin{equation}
\label{81}
\partial_{i}M_{ij}=0.
\end{equation}

As the gauge orbits lay down by the eigenvectors, are vertical to the physical hypersurface defined by the projective metric $M_{ij}(M^2=M)$, the infinitesimal gauge transformation for the velocity potential can be computed as
\begin{equation}
\delta u_i = \partial_i\varepsilon,
\end{equation}
which was also obtained in symplectic context. 

At this stage, we would like to make some comments about the geometry associated to the gauge symmetry. From the Lagrangian (\ref{78}), we identify a singular projective metric which defines a physical surface (orbit space) of the metafluid theory. Since this metric does not depend on the phase space fields, the curvature tensor is null, therefore, the orbit space is flat.

Now, we investigate the metafluid theory described by the Lagrangian (\ref{78}) using the projector method\cite{CP}. From eqn.(\ref{78}), the  canonically conjugated momenta to the field $u_i$ are computed as
\begin{equation}
\label{80}
\pi_{i} = M_{ij}\dot{u}_{j} - {{\partial_{i}n}\over{\partial^{2}}} - \nu M_{ij}\partial^{2}u_{j}.
\end{equation}

Contracting these canonical momenta with the eigenvectors $\partial_{i}$, the primary constraint is determined as
\begin{equation}
\label{84}
\chi \equiv  \partial_{i}\pi_{i}+ n
\end{equation}
which agrees with the results obtained in the symplectic analysis.

From the Lagrangian (\ref{78}), the Euler-Lagrange vector is obtained as
\begin{equation}
\label{85}
E_{k}=M_{kj}\ddot{u}_{j}-c^2 M_{kj}\partial^{2}u_{j}-J_{k}-{{\partial_{k}\dot n}\over{\partial^{2}}}-\nu^{2}M_{kj}\partial^{4}u_{j}.
\end{equation}

Using the projector method\cite{CP}, the equations of motion that satisfy the constraints are projected by the singular metric, namely,
\begin{equation}
\label{86}
M_{ik}E_{k}=0,
\end{equation}
and, then, we get the wave equation for the velocity\cite{Marmanis}, given by,
\begin{eqnarray}
\label{88}
M_{ij}\ddot{u}_{j} -c^2 M_{ij}\partial^{2}u_{j} - M_{ij}J_{j}-\nu^2 M_{ij}\partial^{4}u_{j} &=& 0,\nonumber\\
{\ddot u_{i}}^{\perp} - c^2 \partial^{2}u_{i}^{\perp} - J_{i}^{\perp} - \nu^2  \partial^{4}u_{i}^{\perp} &=& 0,
\end{eqnarray}
where the velocity and current were redefined as $u_{i}^{\perp}=M_{ij}u_{j}$ and  $J_{i}^{\perp}=M_{ij}J_{j}$, respectively,
and $\perp$ denotes the transverse fields (gauge invariant fields). This set of equations together with the constraint condition, eqn.(\ref{84}), define the metafluid dynamics with transverse fields. Due to this, the incompressibility condition is satisfied automatically,
namely,
\begin{eqnarray}
\label{92}
{\partial_i} {u_{i}}^{\perp} &=& \partial_{i}M_{ij}u_{j}\nonumber\\
{\partial_i} {u_{i}}^{\perp} &=& 0.
\end{eqnarray}

Contracting the eigenvector $\partial_{i}$ with the Euler-Lagrange vector, 
\begin{equation}
\label{94}
\partial_{k}E_{k}=\partial_{k}(M_{kj}\ddot{u}_{j})-c^2 \partial_{k}(M_{kj}\partial^{2}u_{j})-
\partial_{k}J_{k}-\partial_{k}{{\partial_{k}\dot n}\over{\partial^{2}}}-\nu^{2}\partial_{k}
(M_{kj}\partial^{4}u_{j}),\nonumber
\end{equation}
we get the continuity equation, given by,
\begin{equation}
\label{95}
\partial_{k}E_{k}=\dot n + \partial_{k}J_{k} = 0.
\end{equation}

As $n = -\nabla^2\phi$, we obtain
\begin{eqnarray}
\label{95a}
- \nabla^2\dot\phi + \partial_{k}J_{k} &=& 0,\nonumber\\
\partial_{k}J_{k} &=& \nabla^2\dot\phi = 0,
\end{eqnarray}
because there is no energy dissipation in the metafluid theory, since it was constructed at the inertial range. Therefore, at this region, the current satisfies the divergenceless condition.

\section{Spectrum of the theory}

The close resemblance of the equations (\ref{04}) to the equations of electromagnetism immediately suggests that turbulent flows have a
hidden wave character in them. Indeed, substituting (\ref{05}) and $\vec w=\nabla \times \vec u$ in the last equations of (\ref{04}), it
can be shown that the average velocity field obeys a wave equation, given by
\begin{equation}
\label{96}
{{\partial^2{\vec u}}\over {\partial t^2}}=c^2 {\nabla^2}{\vec u}+\nu^2{\nabla^4}{\vec u} +{\vec J}_{tr},
\end{equation}
where ${\vec J}_{tr}$ is the turbulent transverse current density.

In the homogeneous turbulence, the turbulent sources do vanish and eqn.(\ref{96}) has the following form
\begin{equation}
\label{97}
{{\partial^2{\vec u}}\over {\partial t^2}}=c^2 {\nabla^2}{\vec u}+\nu^2{\nabla^4}{\vec u} 
\end{equation}
or 
\begin{equation}
\label{98}
{\cal D}{\vec u}=\vec 0
\end{equation}
where the  ${\cal D}$ operator is given by
\begin{equation}
\label{99}
{\cal D}={\partial^{2}_t}-c^2 {\nabla^2}-\nu^2{\nabla^4}.
\end{equation}

Now, we can get the Feynmann propagator or the Green function for eqn.(\ref{98}),
\begin{eqnarray}
iS_{ij}^{F}(x-x^{\prime})&=&\langle 0\vert T(u_{i}(x)u_{j}(x^{\prime}))\vert 0\rangle \label{100}, \nonumber\\
S_{ij}^{F}(x-x^{\prime})&=& -\int {{{\rm d}^{4}{\rm k}}\over{(2\pi)^{4}}}{e^{-i{\rm k}(x-x^{\prime})}\over{k_{0}^{2}-
\omega_{k}^{2}}} \left( \delta_{ij} -{{k_{i}k_{j}}\over{\vec k}} \right),
\label{103}
\end{eqnarray}
with
 
\begin{eqnarray}
\label{104}
T(u_{i}(x)u_{j}(x^{\prime}))&=&\theta(t-t^{\prime})u_{i}(x)u_{j}(x^{\prime})+\theta(t^{\prime}-t)u_{j}(x^{\prime})u_{i}(x),\nonumber\\ 
{\rm k}&=&k_{0}-{\vec k},\nonumber\\
x&=&x_{0}-{\vec x},\\
\omega_{k}^{2}&=& c^{2}{\vec k}^{2}-\nu^{2}{\vec k}^{4},\nonumber
\end{eqnarray}
where the last equation is the dispersion relation to the wave equation (\ref{96})\cite{Marmanis}. 

The calculation of $S_{ij}^{F}(x-x^{\prime})$ can be performed as usual and the result for equal time Feynmann propagator is given by 
\begin{equation}
\label{109}
iS_{ij}^{F}({\vec x}-{\vec {x^{\prime}}})={1\over 2}\int {{{\rm d}^{3}{\vec k}}\over{(2\pi)^{3}}}{{e^{i{\vec k}.({\vec x}-
{\vec {x^{\prime}})}}\over{\omega_{k}}}} \left( \delta_{ij} -{{k_{i}k_{j}}\over{\vec k}} \right).
\end{equation}

Now, summing over $(i,j)$ for $i=j$, we get
\begin{equation}
\label{110}
\sum_{i=j}^{3}iS_{ij}^{F}({\vec x}-{\vec {x^{\prime}}})=\int {{{\rm d}^{3}{\vec k}}\over{(2\pi)^{3}}}{{e^{i{\vec k}.
({\vec x}-{\vec {x^{\prime}})}}\over{\omega_{k}}}}.
\end{equation}

Therefore, eqn.(\ref{110}) can be identified with the two-point correlation function of the velocity field at equal time. So
\begin{equation}
\label{111}
\langle {\vec u}(\vec x).{\vec u}(0)\rangle =\sum_{i=j}^{3}iS_{ij}^{F}({\vec x}-{\vec {x^{\prime}}})=\int {{{\rm d}^{3}{\vec k}}
\over{(2\pi)^{3}}}{{e^{i{\vec k}.({\vec x}-{\vec {x^{\prime}})}}\over{\omega_{k}}}}
\end{equation}
or in the Fourier space, we get
\begin{equation}
\label{112}
\langle {\vec u}(\vec k).{\vec u}(-\vec k)\rangle =\int {\rm d}^{3}{\vec x}\;e^{-i{\vec k}.{\vec x}}\langle {\vec u}(\vec x).
{\vec u}(0)\rangle = \, \omega_{k}^{-1}. 
\end{equation}

With this calculation, we showed $\langle {\vec u}(\vec k).{\vec u}(-\vec k)\rangle \propto E(k)  \propto k^{-1}$. Such result would be expected because of the analogy between electromagnetism and the metafluid dynamics. The reason why in this model we get this exponent instead of the value $-5/3$ found in literature is that the linearization implies the decomposition of the field into various oscillators do not invoke any interaction among them. To get the exponent found in the literature on can proceede as done by Zakharov {\it et al} \cite{Zakharov,Marmanis}, where a perturbation theory was performed. Despite of the value obtained in this formalism, one can see the dispertion relation $\omega(\vec k) $ presents the physics expected for a turbulent flow once, $\omega (\vec k) $ can be real, with positive or negative, or even imaginary. This means in the context of Fourier space that one can have waves with positive or negative energy and that $\omega (\vec k) $ been imaginary gives a dissipative behavior, but that can happen only for sufficiently high wavenumbers. 

\section{Conclusion} 

In this paper, we proposed to investigate incompressible turbulent hydrodynamics in the context of metafluid dynamics in order to open up the possibility to apply all the machinery very well known in quantum field theory. This investigation was possible due to the analogy of incompressible turbulent hydrodynamics with Maxwell electromagnetism. Exploiting this analogy, the Dirac brackets was computed through the symplectic formalism, which revealed a hidden symmetry presents on the metafluid gauge theory, only preserved in inertial range. Afterward, the geometrical meaning of the gauge symmetry was given, showing that the orbit space is flat and, consequently, the divergenceless condition ($\nabla . {\vec J}^{\perp} = 0$) is naturally obtained, at least in the inertial range. Subsequently, the wave equations for the transverse (horizontal) velocity field were computed, which was also obtained in \cite{Marmanis}, but written in terms of the velocity field, both horizontal (gauge invariant) and vertical (gauge variant) components. To finish, the 3D spectrum of metafluid theory was computed, leading to a proper result which does not agree with the usual one computed in the incompressible turbulent hydrodynamics context\cite{spectrum}. It seems that the approximate approach delineated in \cite{Marmanis} and extended here with the introduction of the Lagrangian, given in eqn.(\ref{20}), change the spectrum. We suspect that some singularities due to the nonlinear nature of the incompressible turbulent hydrodynamics are eliminated by the analogy process used to linearize the theory. We believe that the contribution of these singularities in the metafluid theory can appear if the orbit space has a curvature. Despite of not getting the usual exponent for the energy spectrum, we showed that with this formalism, one gets the right behavior for the turbulent flow, in the inertial range.

\section{Acknowledgements} 
This work was partially supported by CNPq and FAPEMIG. Two of us (A.C.R.M and C. N.) thank to CNPq and FAPEMIG (grant no. CEX-00005/00),
respectively, for the financial support and  (W.O and F.I.T.) would like to thank FAPEMIG and CNPq for partial support.

\appendix
\section{Symplectic formalism}

Faddeev and Jackiw \cite{Faddeev-Jackiw} and Barcelos and Wotzasek \cite{Barcelos-Wotzasek} showed how to implement the constraints directly
into the canonical part of the first-order Lagrangian once this method is applied to first
order Lagrangians. In order to systematize the symplectic gauge formalism, a general 
noninvariant mechanical model that has its dynamics governed by a Lagrangian 
${\cal L}(a_i,\dot a_i,\, t)$ (with $i=1,2,...,N$) is considered, where $a_i$ 
and $\dot a_i$ are the space and velocities variables respectively. Notice that this 
consideration does not lead to lost of generality. In the symplectic method, the first-order Lagrangian written in terms of the sympletic variables
 $\xi^{(0)}_{\alpha}(a_i,p_i)$ (with $\alpha =1,2,...,2N$), is required
\begin{equation}
\label{22}
{\cal L}^{(0)}=A^{(0)}_{\alpha}\dot \xi^{(0)}_{\alpha} -U^{(0)},
\end{equation}
where $A^{(0)}_{\alpha}$ is the one-form canonical momenta, $(0)$ indicates that it 
is the zeroth-iterative Lagrangian and, $U^{(0)}$, the sympletic potencial. After that, 
the sympletic tensor, defined as
\begin{equation}
\label{23}
f^{(0)}_{\alpha \beta}={{\partial A^{(0)}_{\beta}}\over 
{\partial \xi^{(0)}_{\alpha}}}-{{\partial A^{(0)}_{\alpha}}\over 
{\partial \xi^{(0)}_{\beta}}},
\end{equation}
is computed. Since this sympletic matrix is singular, it has a zero-mode $(\nu^{(0)})$
 that generates a new constraint when contracted with the gradient of potencial, namely,
\begin{equation}
\label{24}
\Omega^{(0)}=\nu^{(0)}_{\alpha}{{\partial U^{(0)}}\over {\partial \xi^{(0)}_{\alpha}}}.
\end{equation}
Through a Lagrange multiplier $\lambda$, this constraint is introduced into the 
zeroth-iterative Lagrangian (\ref{22}), generating the next one
\begin{eqnarray}
{\cal L}^{(1)}&=&A^{(0)}_{\alpha}\dot \xi^{(0)}_{\alpha} -U^{(0)}+{\dot \lambda}\Omega^{(0)},
\nonumber\\
&=&A^{(1)}_{\alpha}\dot \xi^{(1)}_{\alpha} -U^{(1)}
\label{25}
\end{eqnarray}
where
\begin{eqnarray}
\label{26}
U^{(1)}&=&U^{(0)}\vert_{\Omega^{(0)}=0},\nonumber \\
\xi^{(1)}_{\alpha} &=& (\xi^{(0)}_{\alpha},\lambda),\\
A^{(1)}_{\alpha} &=& A^{(0)}_{\alpha}+\lambda{{\partial \Omega^{(0)}}\over {\partial \xi^{(0)}}}.\nonumber
\end{eqnarray}
The first-iterative sympletic tensor is computed as
\begin{equation}
\label{27}
f^{(1)}_{\alpha \beta}={{\partial A^{(1)}_{\beta}}\over {\partial \xi^{(1)}_{\alpha}}} -
{{\partial A^{(1)}_{\alpha}}\over {\partial \xi^{(1)}_{\beta}}}.
\end{equation}
Since this tensor is nonsingular, the iterative process stops and the Dira´s brackets among the phase-space variables are obtained from
the inverse matrix. On the other hand, if the tensor is singular, a new constraint arises and the iterative process goes on.


\begin{thebibliography}{100}

\bibitem{Landau}L.D. Landau and E.M. Lifshits, {\it Fluid Mechanics} (Pergamon Press, Oxford, 1980).

\bibitem{Monin-Yagolm} A.S. Monin and A.M. Yagolm, {\it Statistical Fluid Mechanics: Mechanics of Turbulence} (The MIT Press, Cambridge, 1971).

\bibitem{Frisch} U. Frisch, {\it Turbulence: the Legacy of A.N. Kolmogorov}, Cambridge University Press, Cambridge (1995).

\bibitem{Bramwell-Holdsworth-Pinton} S.T. Bramwell, P.C.W. Holdsworth and J.F. Pinton, Nature {\bf 396}, 552 (1998).

\bibitem{L'vov} V.S. L'vov, Nature {\bf 396}, 519 (1998).

\bibitem{Periwal} V. Periwal, cond-mat/9602123 (1996).

\bibitem{Polyakov} A.M. Polyakov, Nucl. Phys. {\bf B396}, 367 (1993).

\bibitem{Gurarie} V. Gurarie, {\it Field Theory and the Phenomenon of Turbulence}, (Proc. Recent progress in statistical mechanics and quantum field theory, Los Angeles, USA)(1994).

\bibitem{Eyink-Goldenfeld} G. Eyink and N. Goldenfeld, Phys. Rev. {\bf E50}, 4679 (1994). 

\bibitem{Nelkin} M. Nelkin, Phys. Rev. {\bf A9}, 388 (1974).

\bibitem{deGennes} P.G. De Gennes, {\it Fluctuation, Instability and Phase Transition}, (Proc. NATO Adv. Study Inst., Geilo, Norway). T. Riste, ed. (Noordhoff, Leiden), series B (1975), p1.

\bibitem{Rose-Sulem} H.A. Rose and P.L. Sulem, J. de Phys. {\bf 39}, 441 (1978).

\bibitem{Marmanis} Haralambos Marmanis, Phys. Fluids {\bf 10}, 1428 (1998).

\bibitem{FJ} L. Faddeev and R. Jackiw, Phys. Rev. Lett. {\bf 60}, 1692 (1988);\\
J. Barcelos Neto and C. Wotzasek, Mod. Phys. Lett. {\bf A7}, 1737 (1992); Int. J. Mod. Phys. {\bf A7}, 4981 (1992).

\bibitem{spectrum} L. Ts. Adzhemyan, M. Hnatich, D. Horv\'ath and M. Stehlik, Phys. Rev. {\bf D58}, 4511 (1998).

\bibitem{Jackson} J.D. Jackson, {\it Classical Electrodynamics} (J. Willey, New York, 1983).

\bibitem{Arnold} V.I. Arnold, {\it Mathematical Methods of Classical Mechanics}, (Springer, New York, 1989).

\bibitem{Arnold-Khesin} V.I. Arnold and B.A. Khesin, Ann. Rev. Fluid Mech. {\bf 24}, 145 (1992).

\bibitem{Zeitlin} V. Zeitlin, J. Phys. {\bf A25}, L171 (1992).

\bibitem{Richardson} L.F. Richardson, {\it Weather prediction by numerical process} (Canbridge University Press, Canbridge, 1922).

\bibitem{Itzykson} C. Itzykson and J. B. Zuber, {\it Quantum Field Theory} (McGraw-Hill Inc., New York, 1980).

\bibitem{Henneaux} M. Henneaux and C. Teitelboim, {\it Quantization of Gauge Systems} (Princeton University Press, Princeton, 1992). 

\bibitem{NILL} M. Grabiak, B. Muller and W. Greiner, Ann. Phys. (NY) 172 (1986) 213;\\ 
N. Ilieva, L. Litov, Proceedings, {\it Selected topics in QFT and mathematical physics}, Liblice (1989) 239-250.

\bibitem{CP} C. Amaral e P. Pitanga, Nuovo Cimento {\bf B25}, 817 (1982); Rev. Bras. Fis., {\bf 12} (3), 473 (1982).

\bibitem{Zakharov} Zakharov V.E., L'vov V.S., and Falkovich G., {\it Kolmogorov spectra of turbulence I. Wave Turbulence}, Springer-Verlag, New York, 1992.

\bibitem{Faddeev-Jackiw} L. Faddeev and R. Jackiw, Phys. Rev. Lett. {\bf 60}, 1692 (1988).

\bibitem{Barcelos-Wotzasek} J. Barcelos Neto and C. Wotzasek, Mod. Phys. Lett. {\bf A7}, 1737 (1992); Int. J. Mod. Phys. {\bf A7}, 4981 (1992).

\end{thebibliography}
\end{document}